\documentclass[graybox]{svmult}

\usepackage{mathptmx}       
\usepackage{helvet}         
\usepackage{courier}        
\usepackage{type1cm}        
\usepackage{makeidx}         
\usepackage{graphicx}        
\usepackage{multicol}        
\usepackage[bottom]{footmisc}

\makeindex             

\usepackage{amssymb}

\begin{document}

\title*{Squiral Diffraction}
\author{Uwe Grimm and Michael Baake}
\institute{Uwe Grimm \at 
Department of Mathematics and Statistics, The Open University,
Walton Hall, Milton Keynes\break MK7 6AA, United Kingdom.
\email{u.g.grimm@open.ac.uk}
\and Michael Baake \at 
Fakult\"{a}t f\"{u}r Mathematik, Universit\"{a}t Bielefeld,
Postfach 100131, 33501 Bielefeld, Germany.\break
\email{mbaake@math.uni-bielefeld.de}}
\maketitle

\abstract*{The Thue-Morse system is a paradigm of singular continuous
  diffraction in one dimension. Here, we consider a planar system,
  constructed by a bijective block substitution rule, which is locally
  equivalent to the squiral inflation rule. For balanced weights, its
  diffraction is purely singular continuous. The diffraction measure
  is a two-dimensional Riesz product that can be calculated explicitly.}
\abstract{The Thue-Morse system is a paradigm of singular continuous
  diffraction in one dimension. Here, we consider a planar system,
  constructed by a bijective block substitution rule, which is locally
  equivalent to the squiral inflation rule. For balanced weights, its
  diffraction is purely singular continuous. The diffraction measure
  is a two-dimensional Riesz product that can be calculated explicitly.}

\section{Introduction}

The diffraction of crystallographic structures and of aperiodic
structures based on cut and project sets (or model sets) is well
understood; see \cite{BaaGri-BG11,BaaGri-BG12a} and references
therein. These systems (in the case of model sets under suitable
assumptions on the window) are pure point diffractive, and the
diffraction can be calculated explicitly.

The picture changes for structures with continuous diffraction. Not
much is known in general, in particular for the case of singular
continuous diffraction, even though both absolutely and singular
continuous diffraction show up in real systems
\cite{BaaGri-WW,BaaGri-W}. The paradigm of singular continuous
diffraction is the Thue-Morse chain, which in its balanced form
(constructed via the primitive inflation rule $1\mapsto 1\bar{1}$,
$\bar{1}\mapsto\bar{1}1$ with with weights $1$ and $\bar{1}=-1$, say)
shows purely singular continuous diffraction. This was shown by
Kakutani \cite{BaaGri-Kaku}, see also \cite{BaaGri-BG08}, and the
result can be extended to an entire family of generalised Thue-Morse
sequences \cite{BaaGri-BGG12}.

Here, we describe a two-dimensional system which, in its balanced
form, has purely singular continuous diffraction. For more detail and
mathematical proofs, we refer to \cite{BaaGri-BG12b}. Again, it is
possible to obtain an explicit formula for the diffraction measure in
terms of a Riesz product, with convergence in the vague topology.

\section{The squiral block inflation}

\begin{figure}
\sidecaption
\includegraphics[width=0.35\textwidth]{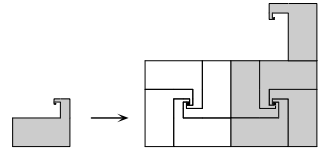}
\caption{The primitive inflation rule for the squiral tiling of the 
Euclidean plane.}
\label{BaaGri-sqinfl}
\end{figure}

The squiral tiling (a name that comprises `square' and `spiral')
was introduced in \cite[Fig.~10.1.4]{BaaGri-GS} as an example of an
inflation tiling with prototiles comprising infinitely many edges. The
inflation rule is shown in Fig.~\ref{BaaGri-sqinfl}; it is compatible
with reflection symmetry, so that the reflected prototile is inflated
accordingly.

\begin{figure}[b]
\includegraphics[width=\textwidth]{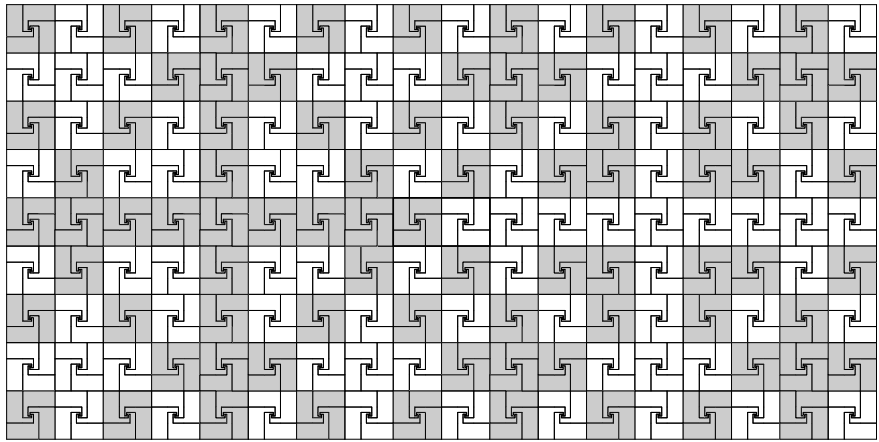}
\caption{Patch of the squiral tiling obtained by two inflation steps
from the central seed.}
\label{BaaGri-sqtil}
\end{figure}

\begin{figure}[t]
\sidecaption
\includegraphics[width=0.5\textwidth]{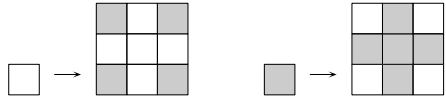}
\caption{Equivalent block inflation rule for the squiral tiling
of Figure~\ref{BaaGri-sqtil}.}
\label{BaaGri-sqblinf}
\end{figure}

A patch of the tiling is shown in Fig.~\ref{BaaGri-sqtil}. Clearly,
the tiling consists of a two-colouring of the square lattice, with
each square comprising four squiral tiles of the same chirality. The
two-colouring can be obtained by the simple block inflation rule sown
in Fig.~\ref{BaaGri-sqblinf}, which is bijective in the sense of
\cite{BaaGri-Nat}. Again, the rule is compatible with colour
exchange. The corresponding hull has $D_{4}$ symmetry, and also
contains an element with exact individual $D_{4}$ symmetry; see
\cite{BaaGri-BG12b} for details and an illustration.

Due to the dihedral symmetry of the inflation tiling, it suffices to
consider a tiling of the positive quadrant. Using the lower left point
of the square as the reference point, the induced block inflation
$\varrho$ produces a two-cycle of configurations $v$ and $\varrho
v$. They satisfy, for all $m,n \ge 0$ and $0 \le r,s \le 2$, the fixed
point equations
\begin{equation}\label{BaaGri-fix}
   (\varrho v)^{}_{3m+r,3n+s} \, = \, 
   \cases{\overline{v}^{}_{m,n}, \quad \rm{if}\;
   r\equiv s\equiv 0 \bmod 2,   \cr 
   v^{}_{m,n},\quad  \rm{otherwise}.  } 
\end{equation}

\section{Autocorrelation and diffraction measure}

For a fixed point tiling under $\varrho^{2}$, we mark each (coloured)
square by a point at its lower left corner $z\in\mathbb{Z}^{2}$. For
the balanced version, each point carries a weight $w^{}_{z}=1$ (for
white) or $w^{}_{z}=\bar{1}=-1$ (for grey). Consider the corresponding
Dirac comb
\begin{equation}\label{BaaGri-def-comb}
  \omega \, = \, w \, \delta^{}_{\mathbb{Z}^{2}} \, = 
  \sum_{z\in\mathbb{Z}^{2}} w^{}_{z} \, \delta^{}_{z}  \, .
\end{equation} 
Following the approach pioneered by Hof \cite{BaaGri-Hof}, the natural
autocorrelation measure $\gamma$ of $\omega$ is defined as\vspace*{-1mm}
\begin{equation}\label{BaaGri-gam-def}\vspace*{-1mm}
   \gamma \, = \, \omega \circledast \widetilde{\omega} \, := \,
   \lim_{N\to\infty} \frac{\bigl(\omega\big|^{}_{C_{N}}\bigr) * 
   \bigl( \widetilde{\omega\big|^{}_{C_{N}}}\bigr)}{(2N+1)^{2}}  \, ,    
\end{equation}
where $C_{N}$ stands for the closed centred square of side length
$2N$.  Here, $\widetilde{\mu}$ denotes the measure defined by
$\widetilde{\mu} (g) = \overline{\mu (\widetilde{g})}$ for $g\in
C_{\mathsf{c}} (\mathbb{R}^{2})$, with $\widetilde{g} (x) :=
\overline{g(-x)}$ (and where the bar denotes complex conjugation).
The autocorrelation measure $\gamma$ is of the form $\gamma =
\eta\delta^{}_{\mathbb{Z}^{2}}$ with autocorrelation coefficients
\begin{equation}\label{BaaGri-eta}
   \eta (m,n) \, = \, \lim_{N\to\infty} \frac{1}{(2N+1)^{2}}
   \sum_{k,\ell=-N}^{N} w^{}_{k,\ell}\, w^{}_{k-m,\ell-n} \, .
\end{equation}
All limits exists due to the unique ergodicity of the underlying 
dynamical system \cite{BaaGri-BG12b}, under the action of the group
$\mathbb{Z}^{2}$.

Clearly, one has $\eta(0,0)=1$, while Eq.~(\ref{BaaGri-fix}) implies the
nine recursion relations
\begin{eqnarray}\label{BaaGri-rec}
  \eta (3m,3n)  \,   & = & \, \eta (m,n)\, , \nonumber \\
  \eta (3m,3n \! + \!1)  \, & = & \, 
   - \textstyle\frac{2}{9} \eta (m,n)  + 
             \frac{1}{3} \eta (m,n \! + \! 1)\, , \nonumber \\
  \eta (3m,3n \! + \! 2)  \, & = & \,   
   \textstyle\frac{1}{3} \eta (m,n)  - 
             \frac{2}{9} \eta (m,n \! + \! 1) \, ,\nonumber \\
  \eta (3m \! + \! 1,3n)  \,   & = & \, 
   - \textstyle\frac{2}{9} \eta (m,n)  + 
             \frac{1}{3} \eta (m \! + \! 1,n) \, ,\nonumber \\
  \eta (3m \! + \! 1,3n \! + \! 1)  \, & = & \, - \textstyle\frac{2}{9} 
        \bigl( \eta (m \! + \! 1,n)  + 
               \eta (m,n \! + \! 1) \bigr) 
               + \frac{1}{9} \eta (m \! + \! 1,n \! + \! 1) \, , \\
  \eta (3m \! + \! 1,3n \! + \! 2)  \, & = & \, - \textstyle\frac{2}{9} 
        \bigl(\eta (m,n)  + 
        \eta (m \! + \! 1,n \! + \! 1) \bigr)
                        + \frac{1}{9} \eta (m \! + \! 1,n) \, ,\nonumber \\
  \eta (3m \! + \! 2,3n)    \, & = & \,   \textstyle\frac{1}{3} 
        \eta (m,n)  - \frac{2}{9} \eta (m \! + \! 1,n)\, ,\nonumber   \\
  \eta (3m \! + \! 2,3n \! + \! 1)  \, & = & \, - \textstyle\frac{2}{9} 
        \bigl(\eta (m,n)  + 
        \eta (m \! + \! 1,n \! + \! 1) \bigr)
                        + \frac{1}{9} \eta (m,n \! + \! 1)\, ,\nonumber   \\
  \eta (3m \! + \! 2,3n \! + \! 2)  \, & = & \,  \textstyle \frac{1}{9} 
       \eta (m,n)  - \frac{2}{9} 
                          \bigl(\eta (m \! + \! 1,n)  + 
       \eta (m,n \! + \! 1) \bigr) ,\nonumber 
\end{eqnarray}
which hold for all $m,n\in\mathbb{Z}$ and determine all coefficients
uniquely \cite{BaaGri-BG12b}. The autocorrelation coefficients
show a number of remarkable properties, which are interesting in their own
right, and useful for explicit calculations.   

Since the support of $\omega$ is the lattice $\mathbb{Z}^{2}$, the
diffraction measure $\widehat{\gamma}$ is $\mathbb{Z}^{2}$-periodic
\cite{BaaGri-B02}, and can thus be written as
\[
   \widehat{\gamma} \, = \, \mu * \delta^{}_{\mathbb{Z}^{2}} \, ,
\]
where $\mu$ is a positive measure on the fundamental domain
$\mathbb{T}^{2}=[0,1)^{2}$ of $\mathbb{Z}^{2}$. One can now analyse
$\widehat{\gamma}$ via the measure $\mu$, which, via the
Herglotz-Bochner theorem, is related to the autocorrelation
coefficients by Fourier transform
\[
    \eta(k) \, = \, \int_{\mathbb{T}^{2}} e^{2\pi \mathrm{i} k z} \, 
                    \mathrm{d}\mu(z)\, ,
\]
where $k=(m,n)\in\mathbb{Z}^{2}$ and $kz$ denotes the scalar product.
We now sketch how to determine the spectral type of $\mu$, and how to
calculate it.

Defining $\Sigma (N) \, := \, \sum_{m,n=0}^{N-1} \eta (m,n)^{2}$, 
the recursions (\ref{BaaGri-rec}) lead to the estimate
\[
   \Sigma (3N) \, \le \, \frac{319}{81}\, \Sigma (N) \, ,
\]
so that $\Sigma(N)/N^{2}\longrightarrow 0$ as $N\to\infty$.  An
application of Wiener's criterion in its multidimensional version
\cite{BaaGri-BG12b,BaaGri-TAO} implies that $\mu$, and hence also the
diffraction measure $\widehat{\gamma}$, is continuous, which means
that it comprises no Bragg peaks at all.

Since $\eta(0,1)=\eta(1,0)=-1/3$, which follows from
Eq.~(\ref{BaaGri-rec}) by a short calculation, the first recurrence
relation implies that $\eta(0,3^{j})=\eta(3^{j},0)=-1/3$ for all
integer $j\ge 0$.  Consequently, the coefficients cannot vanish at
infinity. Due to the linearity of the recursion relations, the
Riemann-Lebesgue lemma implies \cite{BaaGri-BG12b} that $\mu$ cannot
have an absolutely continuous component (relative to Lebesgue
measure). The measure $\mu$, and hence $\widehat{\gamma}$ as well,
must thus be purely singular continuous.

\section{Riesz product representation}

Although the determination of the spectral type of $\widehat{\gamma}$
is based on an abstract argument, the recursion relations
(\ref{BaaGri-rec}) hold the key to an explicit, iterative calculation
of $\mu$ (and hence $\widehat{\gamma}\:$). One defines the distribution
function $F(x,y) := \mu\bigl( [0,x] \times [0,y] \bigr)$ for rectangles
with $0\le x,y < 1$, which is then extended to the positive quadrant
as
\[
   F (x,y) \, = \, \widehat{\gamma} 
    \bigl( [0,x] \times [0,y] \bigr).
\]
This can finally be extended to $\mathbb{R}^{2}$ via
$F(-x,y) = F(x,-y) = - F(x,y)$ and hence $F(-x,-y) = F(x,y)$.  In
particular, one has $F(0,0)=0$ as well as $F(0,y) = F(x,0)=0$, and $F$
is continuous on $\mathbb{R}^{2}$. The latter property is non-trivial,
and follows from the continuity of certain marginals; see
\cite{BaaGri-BG12b} and references therein for details.

\begin{figure}[t]
  \includegraphics[width=0.48\textwidth]{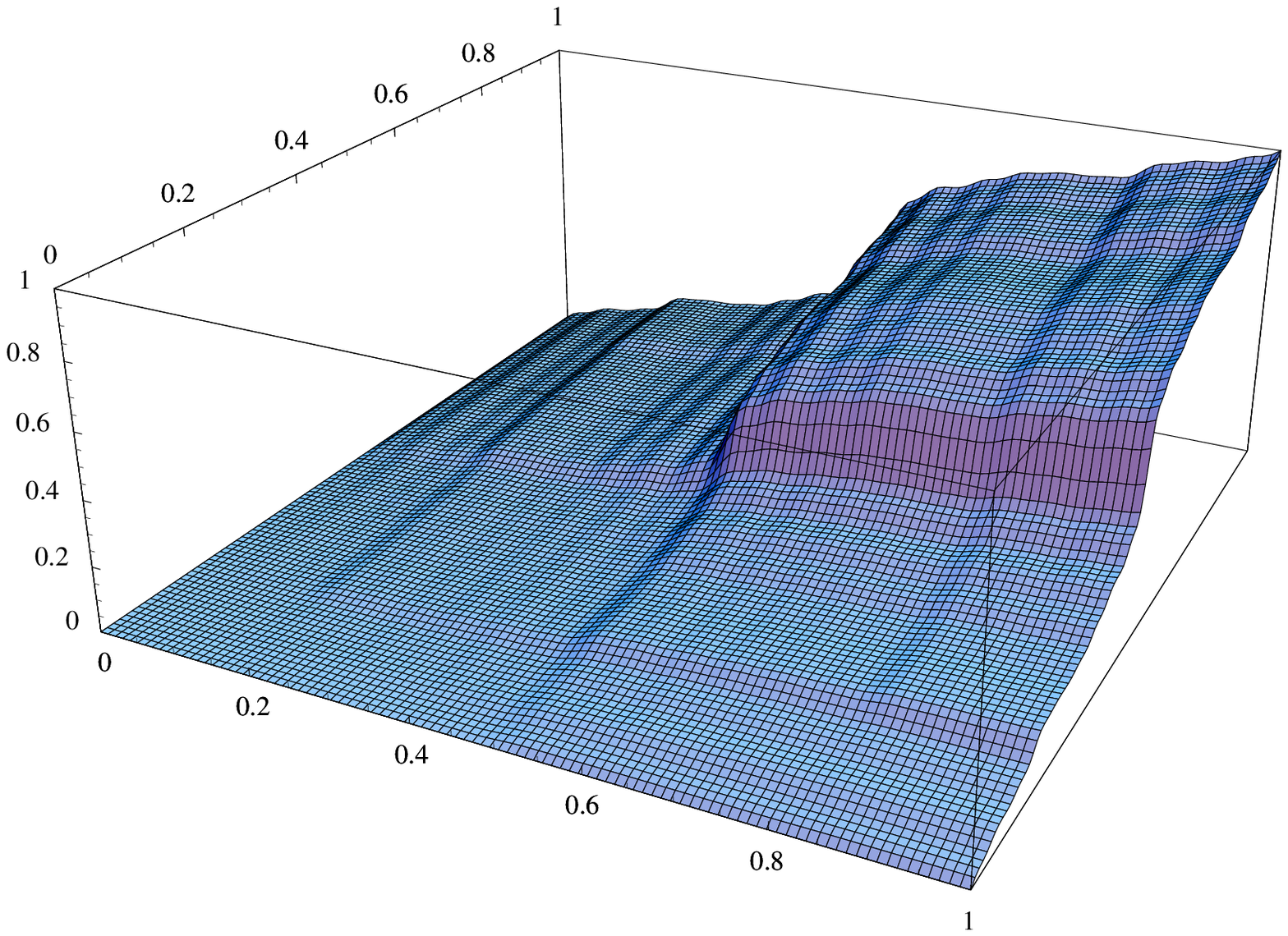}\hspace{\fill}
  \includegraphics[width=0.48\textwidth]{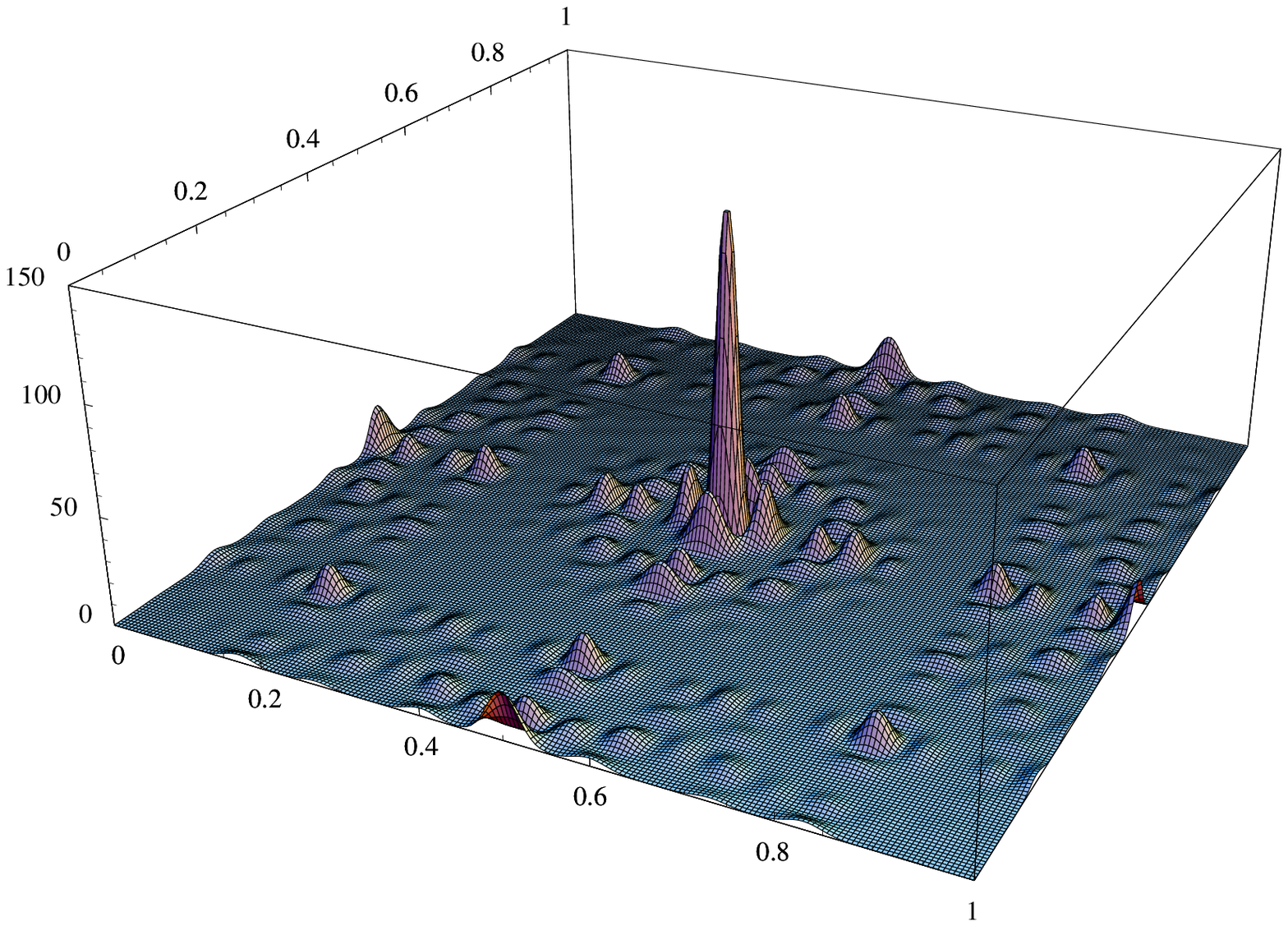}
\caption{The distribution function $F^{(3)}$ 
  of Eq.~(\ref{BaaGri-fiter}) (left) and the corresponding density $f^{(3)}$ 
  of Eq.~(\ref{BaaGri-riesz}) (right), approximating
  the diffraction measure $\widehat{\gamma}$ of the squiral 
  tiling  on $[0,1]^{2}$.}
\label{BaaGri-rfig}
\end{figure}

One can show that, as a result of Eq.~(\ref{BaaGri-rec}), $F$
satisfies the functional relation
\begin{equation}\label{BaaGri-frel}
   F (x,y)  \, = \, \frac{1}{9} \int_{0}^{3x}
      \int_{0}^{3y} \vartheta \bigl( \frac{x}{3}, \frac{y}{3}\bigr)
      \,\mathrm{d} F (x,y)\, ,
\end{equation}
written in Lebesgue-Stieltjes notation, with the 
trigonometric kernel function
\[
   \vartheta (x,y) \, = \,  \frac{1}{9} \bigl(1
   + 2 \cos (2\pi x) + 2 \cos (2\pi y)
   - 4 \cos (2\pi x) \cos (2\pi y)\bigr)^{2} .
\]
The functional relation (\ref{BaaGri-frel}) induces an iterative
approximation of $F$ as follows. Starting from $F^{(0)} (x,y) = xy$
(which corresponds to Lebesgue measure, $\mathrm{d}F^{(0)} = \lambda$) and
continuing with the iteration
\begin{equation}\label{BaaGri-fiter}
   F^{(N+1)} (x,y) \, = \, \frac{1}{9} \int_{0}^{3x}
      \int_{0}^{3y} \vartheta \bigl( \frac{x}{3}, \frac{y}{3}\bigr)
      \,\mathrm{d} F^{(N)} (x,y)\, ,
\end{equation}
one obtains a uniformly (but not absolutely) converging sequence of
distribution functions, each of which represents an absolutely
continuous measure. With
$\mathrm{d}F^{(N)}(x,y)=f^{(N)}(x,y)\,\mathrm{d}x\,\mathrm{d}y$, where
$f^{(N)} (x,y) \, = \, \frac{\partial^{2}}
   {\partial x \ts \partial y} F^{(N)}(x,y)$, one finds the Riesz product
\begin{equation}\label{BaaGri-riesz}
   f^{(N)} (x,y) \, = \prod_{\ell=0}^{N-1} \vartheta (3^{\ell}x,3^{\ell}y)\, .
\end{equation}
These functions of increasing `spikiness' represent a sequence of
(absolutely continuous) measures that converge to the singular
continuous squiral diffraction measure in the vague topology. The case
$N=3$ is illustrated in Figure~\ref{BaaGri-rfig}. Local scaling
properties can be derived from Eq.~(\ref{BaaGri-riesz}).

\section{Summary and outlook}

The example of the squiral tiling demonstrates that the constructive
approach of Refs.~\cite{BaaGri-Kaku,BaaGri-BG08,BaaGri-BGG12} can be
extended to more than one dimension. The result is as expected, and
analogous arguments apply to a large class of binary block
substitutions that are bijective in the sense of
\cite{BaaGri-Nat}. This leads to a better understanding of binary
systems with purely singular continuous diffraction.

It is desirable to extend this type of analysis to substitution
systems with larger alphabets. Although the basic theory is developed
in \cite{BaaGri-Q}, there is a lack of concretely worked-out examples.
Moreover, there are various open questions in this direction,
including the (non-)existence of bijective constant-length substitutions
with absolutely continuous spectrum (the celebrated example from
\cite[Ex.~9.3]{BaaGri-Q} was recently recognised to be inconclusive by
Alan Bartlett and Boris Solomyak).

\begin{acknowledgement}
  We thank Tilmann Gneiting and Daniel Lenz for discussions. This work
  was supported by the German Research Council (DFG), within the CRC
  701.
\end{acknowledgement}

\end{document}